\documentclass[reprint,superscriptaddress,showpacs,amsmath,amssymb,aps,prb]{revtex4-1}
\usepackage{helvet}
\usepackage{graphicx}
\usepackage{hyperref}

\def\SFO{Sr\ensuremath{_3}Fe\ensuremath{_2}O\ensuremath{_{7-\delta}}}
\def\TN{\ensuremath{T_N}}
\def\TD{\ensuremath{T_D}}

\begin{document}

\title{Magnetic phase diagram of \SFO }

\author{D.C. Peets}
\email{dpeets@fkf.mpg.de}
\affiliation{Max-Planck-Institut f\"ur Festk\"orperforschung, D-70569 Stuttgart, Germany}
\affiliation{Department of Physics \&\ Astronomy, University of British Columbia, Vancouver, BC V6T~1Z4, Canada}

\author{J.-H. Kim}
\email{jh.kim@fkf.mpg.de}
\affiliation{Max-Planck-Institut f\"ur Festk\"orperforschung, D-70569 Stuttgart, Germany}

\author{P. Dosanjh}
\affiliation{Department of Physics \&\ Astronomy, University of British Columbia, Vancouver, BC V6T~1Z4, Canada}

\author{M. Reehuis}
\affiliation{Helmholtz-Zentrum Berlin f\"ur Materialien und Energie, D-14109 Berlin, Germany}

\author{A. Maljuk}
\affiliation{Max-Planck-Institut f\"ur Festk\"orperforschung, D-70569 Stuttgart, Germany}
\affiliation{Leibniz Institut f\"ur Festk\"orper- und Werkstoffforschung, D-01171
  Dresden, Germany}

\author{N. Aliouane}
\affiliation{Laboratory for Neutron Scattering, Paul Scherrer Institut, CH-5232 Villigen, Switzerland}

\author{C. Ulrich}
\affiliation{Max-Planck-Institut f\"ur Festk\"orperforschung, D-70569 Stuttgart, Germany}
\affiliation{School of Physics, University of New South Wales, Kensington 2052, Sydney, Australia}
\affiliation{Australian Nuclear Science and Technology Organisation, Lucas Heights, NSW 2234, Australia}

\author{B. Keimer}
\affiliation{Max-Planck-Institut f\"ur Festk\"orperforschung, D-70569 Stuttgart, Germany}

\date{\today}

\begin{abstract}

Magnetometry, electrical transport, and neutron scattering measurements were
performed on single crystals of the Fe$^{4+}$-containing perovskite-related
phase \SFO\ as a function of oxygen content.  Although both the crystal
structure and electron configuration of this compound are closely similar to
those of well-studied ruthenates and manganates, it exhibits very different
physical properties.  The fully-oxygenated compound ($\delta = 0$) exhibits a
charge-disproportionation transition at $\TD = 340$~K, and an
antiferromagnetic transition at $\TN = 115$~K. For temperatures $T \leq \TD$,
the material is a small-gap insulator;  the  antiferromagnetic order is
incommensurate, which implies competing exchange interactions between the
Fe$^{4+}$ moments. The fully-deoxygenated compound ($\delta = 1$) is highly
insulating, and its Fe$^{3+}$ moments exhibit commensurate antiferromagnetic
order below $\TN \sim 600$ K. Compounds with intermediate $\delta$ exhibit
different order with lower $\TN$, likely as a consequence of frustrated
exchange interactions between Fe$^{3+}$ and Fe$^{4+}$ sublattices. A previous
proposal that the magnetic transition temperature reaches zero is not
supported.

\end{abstract}



\maketitle

\section{\label{sec:Intro}Introduction}

Materials in the Ruddlesden-Popper series A$_{n+1}$M$_n$O$_{3n+1}$ with four
$d$-electrons on the transition-metal M-site have proven to be a unique
testbed of correlated-electron physics. Compounds with M = Ru$^{4+}$ (electron
configuration $4d^4$) are generally metallic and display a large variety of
electronic phases including $p$-wave superconductivity (in Sr$_2$RuO$_4$ with
$n=1$), electronic nematicity (in $n=2$ Sr$_3$Ru$_2$O$_7$), and ferromagnetism
(in SrRuO$_3$ with $n=\infty$, the pseudocubic perovskite
structure). Compounds based on Mn$^{3+}$ ($3d^4$), on the other hand, undergo
a variety of orbital and magnetic phase transitions generally leading to
low-temperature insulating states. Iron, which neighbors both Ru and Mn in the
periodic table, can be stabilized in the unusual valence state Fe$^{4+}$ with
$3d^4$ electron configuration in these structures. In contrast to the
ruthenates, Fe$^{4+}$-based perovskites can exhibit either metallic or
correlation-driven insulating states, depending on subtle details of their
crystal structures. For instance, SrFeO$_{3}$ (with $n=\infty$) is metallic at
all temperatures, whereas CaFeO$_{3}$ is charge-disproportionated and
insulating below $T \sim 290$ K.\cite{Takano1977,Takano1983} Neither of these
compounds shows orbital order, which is a common feature of the
Mn$^{3+}$-based manganates, and their magnetic ground states exhibit helical
order,\cite{Takeda1972,Woodward2000,Reehuis2012} which has been observed in
neither the ruthenates nor the manganates. In terms of the itineracy of the
valence electrons, the ferrates therefore fall between the manganates and
ruthenates with $d^4$ electron configuration, yet their phase behavior is
distinctly different from those of both neighbors in the periodic table.

The present study of the $n=2$ Ruddlesden-Popper compound \SFO, the
$3d$-electron analog of Sr$_3$Ru$_2$O$_7$, was motivated by the desire to
obtain further insight into the origin of the pronounced differences between
ruthenates, manganates, and ferrates, and into the influence of the oxygen
stoichiometry on the magnetic phase behavior of the ferrates. Although the
layered crystal structure of \SFO\ is expected to give rise to
quasi-two-dimensional electronic properties, current knowledge of its
properties derives entirely from measurements on powder
samples. M{\"o}{\ss}bauer spectroscopy experiments on fully-oxygenated
Sr$_3$Fe$_2$O$_{7}$ powders indicated disproportionation of Fe$^{4+}$ into the
nominal valence states Fe$^{3+}$ and Fe$^{5+}$ around 340~K,
\cite{Adler1997,Kuzushita2000} closely analogous to the $n=\infty$ analog
CaFeO$_3$. \cite{Takano1977} Magnetic susceptibility measurements at this
doping have indicated an antiferromagnetic transition with N\'eel temperature
$T_N \sim 120$ K,\cite{Kuzushita2000} but neutron powder diffraction
measurements only revealed weak, broad Bragg reflections indicative of
two-dimensional magnetic order,\cite{Dann1993} and these could not be indexed,
so the nature of this transition remains
unclear. It has also been reported that \SFO\ exhibits a continuous
insulator-metal transition under pressure, \cite{Rozenberg1999} likely as a
consequence of changes in the valence-electron bandwidth. \cite{Abbate2004}

One tunable parameter available in these ferrates is oxygen content --- both
\SFO\ and SrFeO$_{3-x}$ can be readily reduced from Fe$^{4+}$ to Fe$^{3+}$.
Since both electronic and oxygen conductivities are high at operating
temperatures, both materials have been investigated as mixed conductors for
solid oxide fuel cell applications,
\cite{Huang2001,Ralph2001,Wincewicz2004,Mogni2005,Prado2007} thus much is
known about the oxygen stoichiometry.  SrFeO$_{3-x}$ supports a series of
oxygen vacancy-ordered phases,~\cite{Mizusaki1992,Hodges2000} separated by
phase-separation regions which extend as high as 900$^\circ$C, complicating
analysis.  Even a
small concentration of oxygen defects in SrFeO$_{3-x}$ can induce a wide
variety of charge-ordered and magnetically-ordered states at low temperatures,
which are associated with unusual magneto-transport phenomena.
\cite{Lebon2004,Adler2006,Ishiwata2011,Reehuis2012}.

\begin{figure}[htb]
\includegraphics[width=\columnwidth]{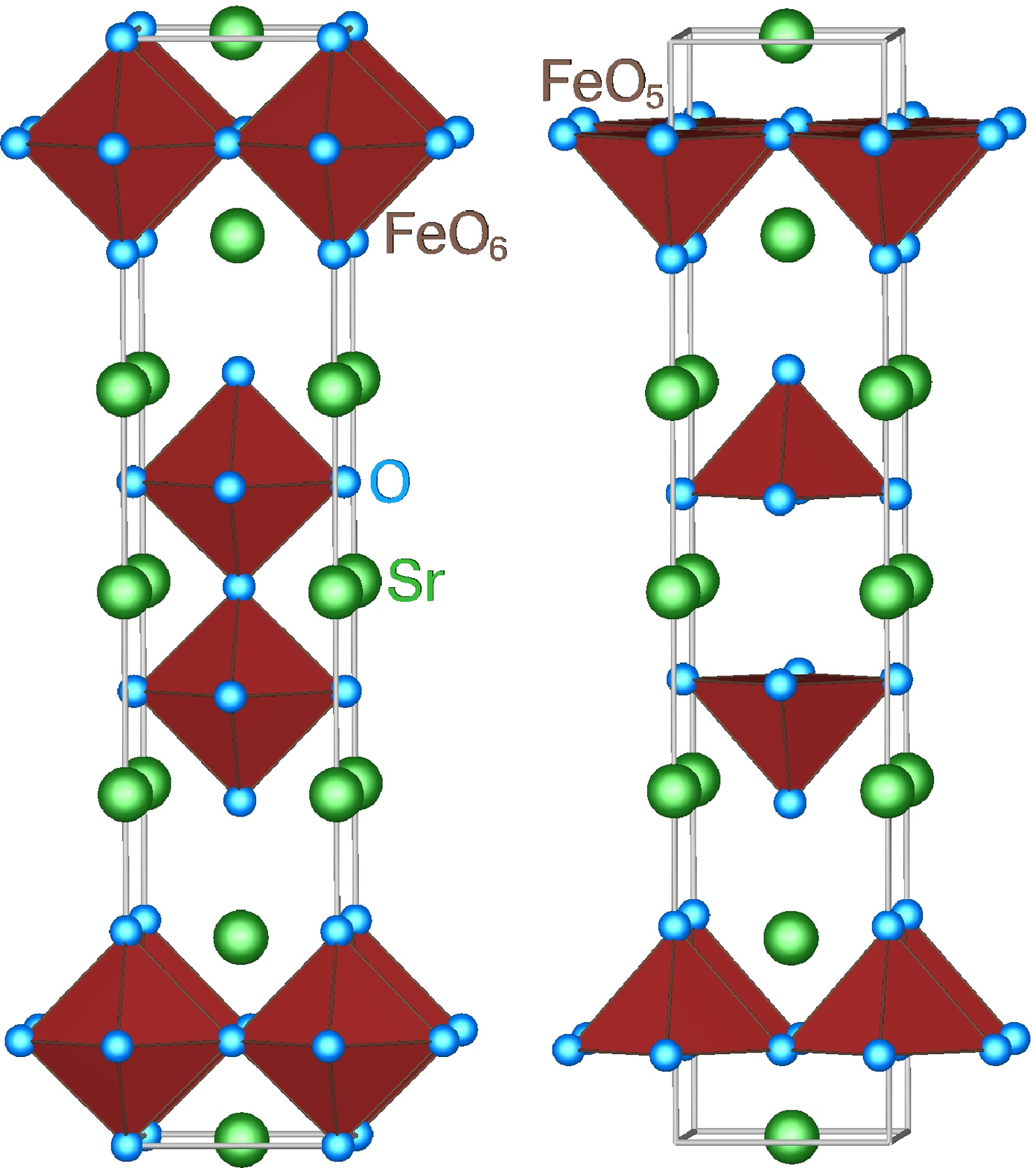}
\caption{\label{fig:structure}(color online) Crystal structures of
  fully-oxygenated Sr$_3$Fe$_2$O$_{7}$ (left) and fully-deoxygenated
  Sr$_3$Fe$_2$O$_{6}$ (right), in which the bridging oxygen site has been
  completely emptied.  Based on refinements in Ref.~\onlinecite{Dann1992}.}
\end{figure}

No evidence for oxygen order has been reported in \SFO. Limited data on \SFO\
powder samples suggest that the antiferromagnetic transition temperature \TN\
in this material decreases as oxygen is removed, possibly approaching zero
near $\delta=0.5$~\cite{Dann1992,MacChesney1966} before recovering for
$\delta=1$, where a magnetic transition with \TN\ as high as 550~K has been
claimed. \cite{Dann1993} This observation would indicate enhanced magnetic
fluctuations in compounds with iron valence intermediate between Fe$^{3+}$
(Sr$_3$Fe$_2$O$_{6}$) and Fe$^{4+}$ (Sr$_3$Fe$_2$O$_{7}$), whose origin has
yet to be determined, and could suggest the presence of a metamagnetic quantum
critical point. Analogy to SrFeO$_{3-x}$ would indicate that deoxygenation
tunes the balance between antiferromagnetic and ferromagnetic coupling,
possibly leading to multiple magnetic ground states, while the lower
dimensionality should lead to narrower bandwidth and more insulating
behavior.\cite{Abbate2004}  Indeed, unlike SrFeO$_{3.00}$,
Sr$_3$Fe$_2$O$_{7.00}$ is insulating at ambient pressure.  Since the oxygen
site emptied on deoxygenation is the bridging site within the bilayer
(Fig.~\ref{fig:structure} contrasts the crystal structures of
Sr$_3$Fe$_2$O$_{6}$ and Sr$_3$Fe$_2$O$_{7}$), doping should alter the
intra-bilayer coupling, effective dimensionality, bandwidth, and bond angles
in addition to the oxidation state of iron, making detailed predictions of the
phase diagram challenging.  The effect of doping in this material has not been
explored in detail, but this information will be crucial to underpin future
studies of the low-temperature behavior.

Motivated by these considerations, we have undertaken a comprehensive
investigation of the magnetic order and electrical conductivity of \SFO\
single crystals with carefully controlled oxygen stoichiometry.  This paper
reports the magnetic transitions observed in magnetization, resistivity, and
simple neutron diffraction measurements, leading to a proposed doping phase
diagram for \SFO.

\section{Experimental details}

Large single crystals of \SFO\ were grown by the floating zone technique under
2.5~atm of oxygen in a four-mirror image furnace, using a growth rate of about
2~mm/h for low mosaicity, as described elsewhere.~\cite{Maljuk2004}  A variety
of different oxygen contents were prepared by annealing for a long enough time
that the samples would fully equilibrate under various temperatures and oxygen
partial pressures, based on the established oxygen phase
diagram. \cite{Mogni2005} With a few exceptions, the anneals were concluded by
quenching the crystals to liquid nitrogen.  The $7-\delta=7.00$ end member was
produced by annealing under 6000~atm of oxygen at 550$^\circ$C for four days,
then cooling to room temperature over the course of a day to maximize the
oxygen content.  The $7-\delta=6.00$ end member was annealed at 650$^{\circ}$C
under 1~atm of 5\% hydrogen in argon and nearby $7-\delta=6.05$ in
argon. Quenching these samples would have required exposing them to air, at
the risk of changing their oxygen contents;  they were thus allowed to cool
freely in the furnace.  It was also impossible to rapidly quench a sample
annealed to an oxygen content of $\sim$6.90, since this was performed under
high oxygen pressure.  In this case, the split-tube furnace was fully opened
and the pressure vessel containing the sample was removed from the heated zone
of the furnace to the extent possible.  Annealing conditions are summarized in
Table~\ref{tab:anneal}.  All annealed and unannealed crystals were stored in
an argon glovebox having less than 0.1~ppm H$_2$O, to minimize the
intercalation of water which causes delamination of the crystal and introduces
transverse cracks.

\begin{table}[htb]
\caption{\label{tab:anneal}Annealing conditions (temperature, oxygen partial
  pressure, and time) used to generate the oxygen contents studied here, and
  approximate oxygen content determined by thermogravimetric (TG) analysis,
  estimated to be accurate to $\pm0.06$~oxygens per formula unit.  Oxygen
  contents $7-\delta$ with standard deviations represent preliminary
  refinement results from single-crystal x-ray (6.50) or neutron diffraction,
  using the published $I4/mmm$ cell; otherwise, the first column reports
  nominal oxygen contents.  The highest-doped sample is referred to as 7.00 in
  this paper. }
\begin{tabular}{|l|r|c|r|c|}\hline
$7-\delta$ & $T$ ($^\circ$C) & $P_{O_2}$ & Time & TG \\ \hline\hline
6.00 & 650 & 0.05 atm H$_2$ in Ar & 7 days & 6.04 \\ \hline
6.05(2) & 880 & 1~atm Ar, O$_2<1$~ppm & 5 days & --- \\ \hline
6.125 & 690 & 0.0001~atm & 6 days & 6.17\\ \hline
6.25 & 665 & 0.002~atm & 6 days & 6.30 \\ \hline
6.33 & 547 & 0.001~atm & 7 days & 6.35 \\ \hline
6.40 & 636 & 0.05~atm & 6 days & 6.41 \\ \hline
6.50(3) & 645 & 1~atm & 6 days & 6.54 \\ \hline
6.67 & 425 & 1~atm & 11 days & 6.73 \\ \hline
6.75 & 350 & 1~atm & 14 days & 6.84 \\ \hline
6.775 & 275 & 1~atm & 15 days & 6.75 \\ \hline
6.90(2) & 350 & 60~atm & 13 days & 6.98 \\ \hline
6.98(2) & 550 & 6000~atm & 4 days & 6.95 \\ \hline
\end{tabular}
\end{table}

The oxygen contents were verified by thermogravimetric analysis in a Netsch
STA-449C DTA/TG apparatus, by heating a portion of each anneal batch in
flowing argon (oxygen partial pressure $\lesssim10^{-7}$~atm) to
1420$^\circ$C, under which conditions the oxygen stoichiometry should be very
close to Sr$_3$Fe$_2$O$_{6.00}$, while monitoring the mass loss. The measured
oxygen contents generally agreed with expectations within the estimated
uncertainty of $\sim$0.06 of an oxygen per formula unit.  Within this error,
the oxygen contents were universally found to be somewhat higher than
expected, perhaps due to additional mass losses from intercalated
water. Samples that were exposed to air for longer periods of time generally
had higher apparent oxygen contents.  In some cases, the water deintercalated
around $180-200^\circ$C, and could be separated from the effects of
deoxygenation, which was too slow to be observable below $\sim350-400^\circ$C
on the timescale of this measurement.  Several oxygen contents were
independently confirmed via preliminary x-ray or neutron diffraction structure
refinements, and these results consistently agreed very well with the intended
oxygen contents.\cite{Reehuis2014} Given the uncertainty in all experimental
results and their consistency with expectations, samples are referred to in
this work by their nominal oxygen contents.

Magnetic measurements were performed on field-cooling by vibrating-sample
magnetometry (Quantum Design MPMS-VSM or Quantum Design PPMS with VSM option)
in fields of 2000~Oe and in some cases additional fields, using quartz sample
holders.  All samples were first aligned by x-ray Laue  diffraction, to ensure
the field was applied along the crystal's tetragonal axes --- for lack of
evidence suggesting a reduction in
symmetry,\cite{Rozenberg1999,Adler1999,Prado2001,Shilova2002} the published
$I4/mmm$ tetragonal unit cell \cite{Brisi1961,Dann1992} is used throughout
this paper.  Magnetization measurements were supplemented at several doping
levels by resistivity, which was measured in a Quantum Design PPMS.  Gold
wires were attached to corners of each thin, $\sim 1$~mm$^2$ sample using
silver epoxy, which was allowed to cure for several hours at 180--200$^\circ$C
in air before the crystal was mounted to the sample puck with GE Varnish (any
stress due to thermal expansion has been neglected).  At these cure
temperatures, the oxygen mobility remains low and the oxygen content should
not change, although this could promote oxygen order.  Due to the crystals'
tendency to delaminate (and often fracture while doing so) when exposed to
air, absolute bulk resistivities cannot be reliably obtained in this
manner. In the following, each sample's resistivity is therefore referenced to
its 300~K zero-field value.  However, it is worth noting that a significant
and monotonic increase in resistivity was observed with oxygen removal, and
the temperature below which the samples became too resistive to measure
increased markedly as the oxygen content was reduced.

To clarify the origin of the phase transitions revealed by the magnetization
measurements, the temperature dependence of selected magnetic Bragg peaks was
measured at several doping levels by neutron diffraction. The measurements
were performed at the Morpheus and TriCS beamlines at SINQ at the
Paul-Scherrer-Institut in Villigen, Switzerland, and on the E5 diffractometer
at the BER-II reactor at the Helmholtz-Zentrum Berlin, Germany.  Full datasets
were in general not collected, as the intention was to determine a phase
diagram to focus future work, rather than perform refinements.  Samples were
typically half-cylinders of $\sim8$~mm diameter and $\sim$5-7~mm in length,
although at a few doping levels rods up to 23~mm in length were used.  Neutron
wavelengths of 2.317\AA\ (TriCS), 4.99\AA\ (Morpheus) and 2.38\AA\ (E5) were
selected using the (002) reflection from pyrolytic graphite (PG)
monochromators, and higher-order contamination ($\lambda/2$) was prevented
through the use of a PG (TriCS, E5) or liquid-nitrogen-cooled beryllium powder
(Morpheus) filter.  Data were collected at TriCS and Morpheus using a point
detector (a 2-inch diameter $^3$He tube), and at E5 using a position-sensitive
$^3$He detector of dimensions $90\times90$~mm$^2$.  Samples were mounted in
four-circle geometry on a closed-cycle refrigerator, and collimators and slits
were set such that each sample was fully illuminated.  For most doping levels,
selected magnetic peaks at each temperature were fit to Gaussian lineshapes,
which were then integrated.  For crystals with oxygen contents $7-\delta$ of
6.05, 6.90, and 7.00, data were instead integrated using the RACER program,
\cite{Wilkinson1988} which uses the parameters describing the shape of strong
peaks to improve the precision in the integration of weaker ones, minimizing
the integral's relative standard deviation.

\section{Results and Discussion}

\begin{figure*}[htb]
\includegraphics[width=\textwidth]{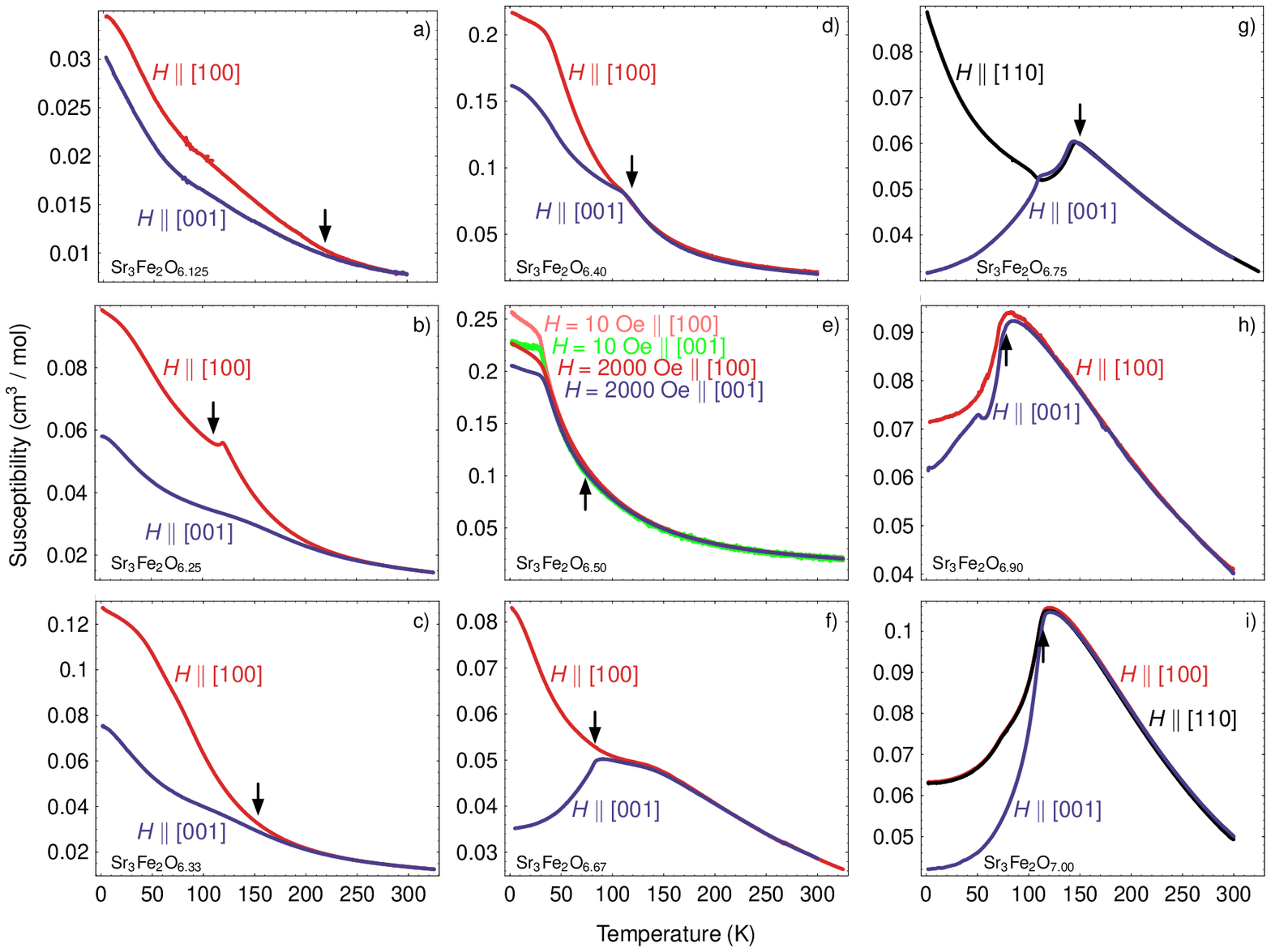}
\caption{\label{fig:mag}(color online) Molar magnetic susceptibility in fields
  of 2000~Oe at a variety of dopings.  The low-temperature susceptibility is
  lower for fields $H \parallel [001]$.  Arrows mark the N\'eel transitions
  found through neutron diffraction at each doping (or at $7-\delta =6.775$ in
  the case of 6.75).}
\end{figure*}

\begin{figure}[htb]
\includegraphics[width=\columnwidth]{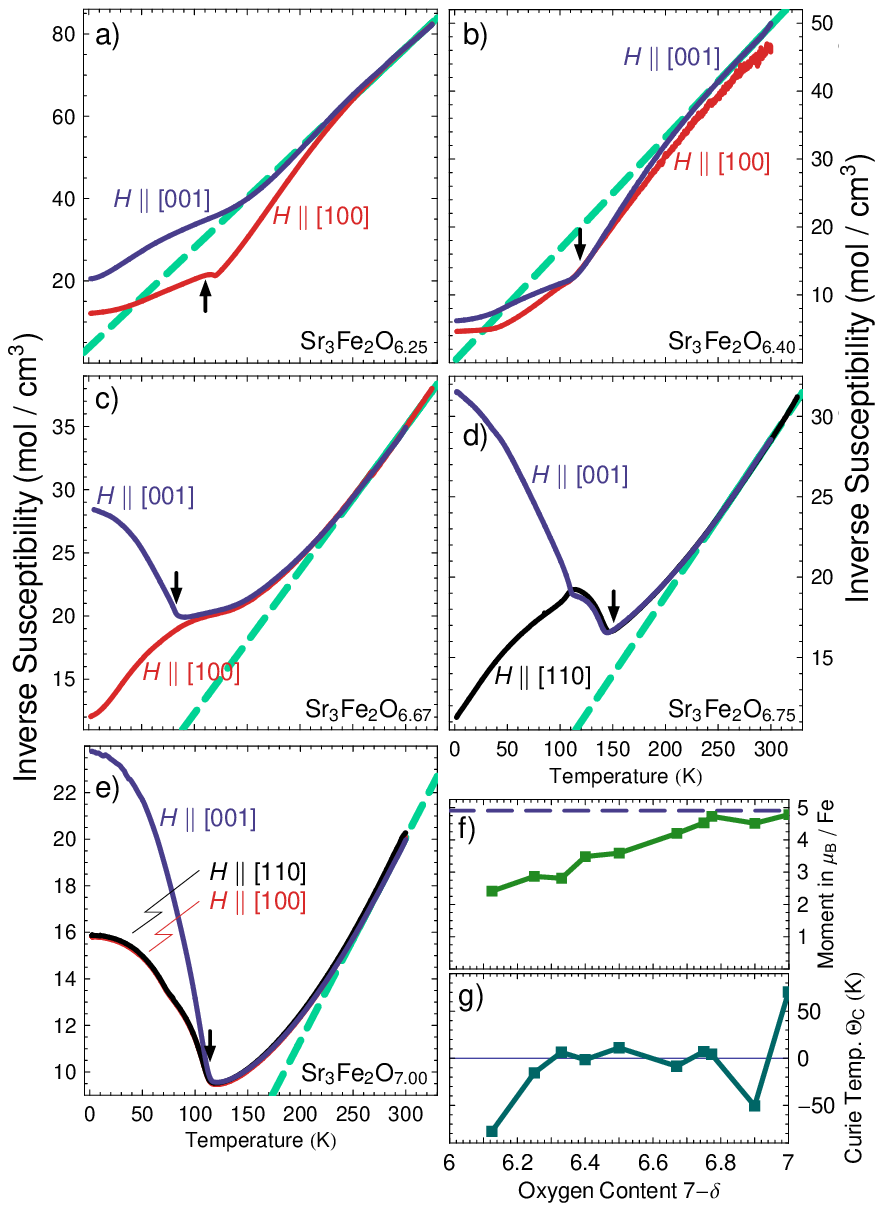}
\caption{\label{fig:curie}(color online) a)--e) Examples of $1/\chi$ vs $T$
  plots, for selected dopings of \SFO:  At temperatures above the highest
  magnetic transition the plots are generally nonlinear, suggesting an onset of
  quasi-two-dimensional spin correlations.  The N\'eel transition from neutron
  diffraction is again marked.  f) and g) Doping dependence of the
  paramagnetic moment and Curie temperature $\Theta_C$ resulting from fits to
  a Curie-Weiss law.  The dashed line indicates the moment
  $g\sqrt{S(S+1)}\mu_B\approx 4.9\mu_B$ expected for high-spin Fe$^{4+}$
  ($S=2$).}
\end{figure}

Field-cooled molar magnetic susceptibility ($\chi$) data at several
representative oxygen concentrations are compiled in Fig.~\ref{fig:mag}.  For
every sample, $\chi$ was measured under 2000~Oe applied fields $H\parallel
[001]$, and $H\parallel [100]$ and/or $[110]$, while at most doping levels
additional data were collected in lower fields (10 or 100~Oe) to verify that
2000~Oe was not excessive. Data were taken for both in-plane directions for
four dopings (shown for $7-\delta=7.00$), but the in-plane field orientations
produced indistinguishable results as would be expected for any effectively
tetragonal material in low fields and in the absence of field-training.  On
other samples only one in-plane field orientation was measured.  Data taken in
2000~Oe were fully consistent with those in lower fields except for minor
differences in Sr$_3$Fe$_2$O$_{6.50}$ (10~Oe data shown), where the higher
field induced saturation at a slightly lower susceptibility (higher
temperature). The high-temperature susceptibilities were fit to a Curie-Weiss
form; however, plots of $1/\chi$ versus $T$ exhibit significant curvature well
above the ordering temperature (Fig.~\ref{fig:curie}). The temperature range
for the Curie-Weiss fits was therefore restricted to $T \geq 250$~K for all
dopings.

For Sr$_3$Fe$_2$O$_{7-\delta}$ with $\delta \leq 0.2$, the paramagnetic 
moment per Fe ion resulting from these fits, $4.8 \mu_B$, is consistent 
with that expected for high-spin Fe$^{4+}$, indicating that the spins 
are weakly correlated at room temperature. The curvature in the $1/\chi$ 
versus $T$ plot at lower $T$, but well above \TN, can be attributed to 
the onset of spin correlations. The layered, body-centered crystal 
structure of \SFO\ (Fig.~\ref{fig:structure}) will give rise to weak and 
frustrated antiferromagnetic interactions between adjacent bilayers. The 
iron moments are therefore expected to be much more strongly coupled 
in-plane than between bilayers, so that the spin correlations close to 
$T_N$ are presumably quasi-two-dimensional.

With decreasing oxygen content, the paramagnetic moment per Fe ion extracted
from these fits decreases, in qualitative agreement with the gradual reduction
of the electron density on the iron ions. However, this reduction is more
pronounced than expected for mixtures of high-spin Fe$^{3+}$ and Fe$^{4+}$
ions according to the oxygen stoichiometry, especially for samples with
$7-\delta \leq 6.40$. In view of the rapidly increasing N\'eel temperature in
this doping range and the expected reduction in the interlayer exchange
coupling as bridging oxygens are removed, we tentatively attribute this
observation to strong quasi-two-dimensional antiferromagnetic spin
correlations in the temperature range where the Curie-Weiss fits were carried
out. Measurements at higher temperatures (under inert-gas atmosphere where
necessary to prevent reoxygenation) will be required to access the
uncorrelated paramagnetic regime at lower oxygen contents.

Because of the narrow temperature range employed for the Curie-Weiss fits, the
resulting paramagnetic Curie-Weiss temperatures ($\Theta_C$) carry large
systematic errors. It is nonetheless notable that their absolute values are
generally below the magnetic ordering temperatures, probably as a consequence
of highly anisotropic exchange interactions.

The susceptibility of every sample with oxygen content $7-\delta > 6.00$
exhibits at least one anomaly indicating a phase transition.   Anisotropic
magnetization typically appeared below one such transition,  clearly
identifying it as magnetic in origin.  In particular, the  susceptibility
clearly indicates an onset of anisotropy at the known antiferromagnetic
transition ($T_N = 115$ K) for the end member Sr$_3$Fe$_2$O$_{7.00}$. At some
lower  oxygen contents the anisotropy is weak but still recognizable (e.g.,
Sr$_3$Fe$_2$O$_{6.50}$ in Fig.~\ref{fig:mag}e). Our observation of magnetic
transitions at all dopings thus does not  confirm conclusions from early
powder data according to which the N\'eel  temperature vanishes for $\delta =
0.50$.~\cite{MacChesney1966} For all  oxygen contents $7-\delta > 6.00$, the
susceptibility is lower for  fields along the $c$-axis than for in-plane
fields, suggesting a spin  orientation with more out-of-plane than in-plane
character.  Susceptibility upturns corresponding to net moments of order
0.01~$\mu_B$ per iron are observed for some doping levels and field
orientations upon cooling below $T_N$.  This may indicate weak ferromagnetic
contributions to the order parameter, perhaps reflecting spin canting due to
Dzyaloshinskii-Moriya interactions. Weakly correlated spins at defect sites of
the crystal lattice probably contribute to the overall  increase of the
susceptibility upon cooling observed for large $\delta$.

For most samples, two anomalies are visible in the magnetization data. For
instance, Figure \ref{fig:mag} shows clear transitions below the
onset of anisotropy in Sr$_3$Fe$_2$O$_{6.25}$ and above it in
Sr$_3$Fe$_2$O$_{6.75}$. A detailed analysis of the structural phase
composition is required to elucidate the origin of the secondary anomalies for
samples with lower oxygen contents. Possible candidates include charge or
orbital order, metamagnetic transitions, and transitions in minority phases
having different oxygen vacancy order, if such order exists in this
compound. In particular, a weak secondary anomaly below $T_N$ in
Sr$_3$Fe$_2$O$_{7.00}$ resembles a transition observed in fully-oxygenated
SrFeO$_{3.00}$ which has been attributed to magnetic order nucleated by a
dilute concentration of residual oxygen
vacancies.\cite{Lebon2004,Adler2006,Reehuis2012}

\begin{figure}[htb]
\includegraphics[width=\columnwidth]{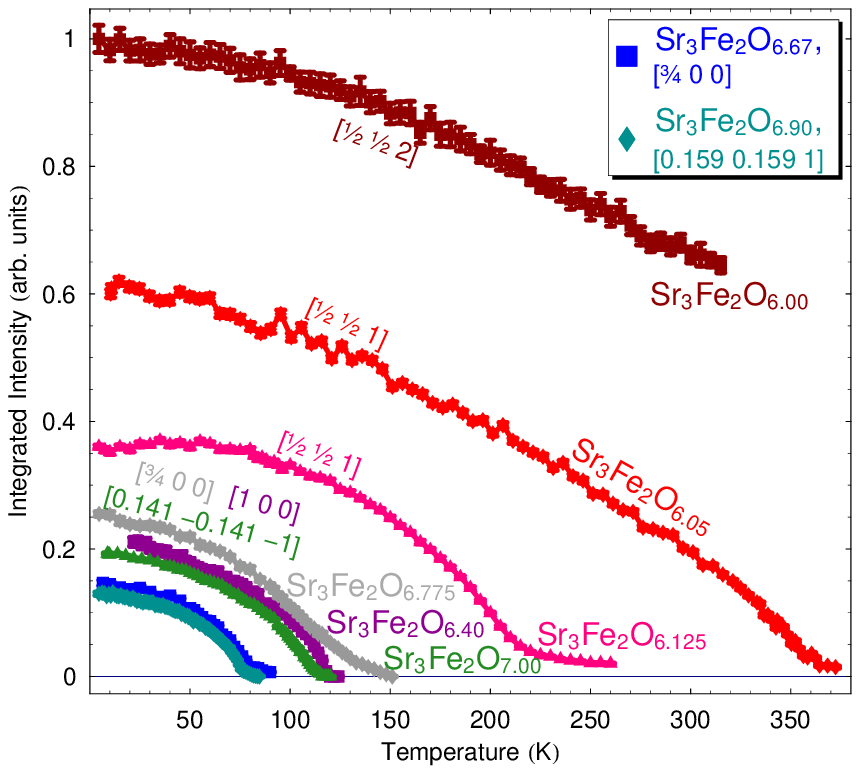}
\caption{\label{fig:neutrontemp}(color online) Temperature dependence of
  integrated intensity in selected magnetic Bragg peaks from neutron
  diffraction, for several oxygen contents $7-\delta$.  Note that these are
  not propagation vectors.  For clarity, the intensity for $T\rightarrow 0$
  has been scaled by the N\'eel temperature. }
\end{figure}

To determine the nature of the antiferromagnetic order, the N\'eel temperature, and the temperature
dependence of the order parameter, single crystal neutron diffraction
experiments were performed at selected doping levels. These experiments were
generally limited to a survey of the first Brillouin zone in reciprocal space
and temperature-dependent measurements of the peaks thus identified, and are
not sufficient to allow determination of a propagation vector or spin
arrangement.  For $7-\delta = 7.00$ and 6.90, surveys of reciprocal space
revealed incommensurate magnetic reflections with wave vectors $[\xi~\xi~L]$
with  $\xi = 0.141$ and 0.159, respectively, and $L =$~integer.  (The wave
vector coordinates are given in reciprocal lattice units based on the
tetragonal space group $I4/mmm$ with lattice parameters $a = b = 3.846$~{\AA}
and $c = 20.234$~{\AA} for $T= 390$~K at $7-\delta = 7.00$.) The temperature
dependence of the magnetic Bragg intensity (shown in
Fig.~\ref{fig:neutrontemp} for selected dopings) indicates second-order phase
transitions with  N\'eel temperatures 114 and 78~K, coincident with the onset
of the anisotropy in the macroscopic susceptibility
(Fig.~\ref{fig:mag}). These observations are analogous to the behavior of the
incommensurate magnetic reflections with wave vector $[\xi~\xi~\xi]$ in the
pseudocubic perovskites SrFeO$_3$ and CaFeO$_3$, which have been attributed to
helical magnetic order.\cite{Takeda1972,Woodward2000,Reehuis2012} It is
therefore quite likely that the magnetic ordering pattern in \SFO\ for small
$\delta$ is an analogous helical state, but hosted in a two-dimensional
system. A detailed crystallographic refinement of the crystal and magnetic
structure is required to confirm this hypothesis.

\begin{figure}[htb]
\includegraphics[width=\columnwidth]{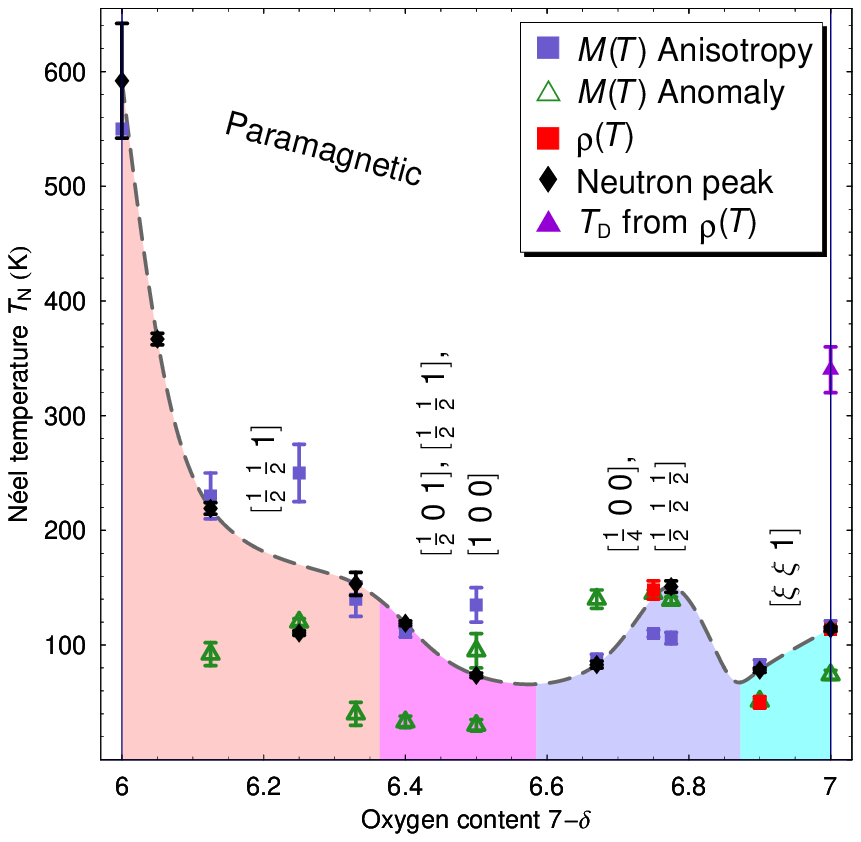}
\caption{\label{fig:TN}(color online) Magnetic doping phase diagram of \SFO,
  constructed from magnetization, resistivity and neutron diffraction
  measurements, including the charge disproportionation temperature \TD.
  Shading is added to indicate where magnetic peaks were found
  at consistent wavevectors; the dashed line respresenting \TN\ is constructed
  primarily based on neutron results and serves as a guide to the eye.
  Note that these wave vectors represent peak positions, not propagation
  vectors.  The neutron diffraction transition for fully deoxygenated
  Sr$_3$Fe$_2$O$_{6.00}$ is an extrapolation as described in the text, and the
  value quoted in Ref.~\onlinecite{Dann1993} for this doping is also
  included.}
\end{figure}

For $7-\delta \leq 6.33$, the neutron diffraction data revealed commensurate
magnetic reflections with wave vectors $[0.5~0.5~L]$, $L$ an integer,
consistent with an earlier report on Sr$_3$Fe$_2$O$_6$ powder.\cite{Dann1993}
The temperature-dependent Bragg intensity again indicates second-order phase
transitions with strongly doping-dependent $T_N$. We were unable to access the
transition temperature for Sr$_3$Fe$_2$O$_{6.00}$ --- observing this
transition without altering the doping will require performing measurements at
high temperatures in the presence of hydrogen --- so its transition
temperature was estimated by scaling the data to match those taken on
Sr$_3$Fe$_2$O$_{6.05}$.  The resulting extrapolated $T_N$ is $600\pm50$~K.
The N\'eel transition for fully deoxygenated Sr$_3$Fe$_2$O$_{6.00}$ has been
quoted as 550~K~\cite{Dann1993} or 298~K~\cite{Dann1992}.  Our neutron results
support the former value. The onset of the  $[0.5~0.5~L]$ intensity in the
samples with $7-\delta = 6.125$ and 6.33 coincides approximately with the
onset of anisotropy in the macroscopic susceptibility (Fig.~\ref{fig:TN}),
confirming the assignment of these anomalies to antiferromagnetic ordering.

In crystals with oxygen contents $7-\delta = 6.40$ and 6.50, peaks were found
at $[0.5~0.5~L]$ and $[0.5~0~L]$ for integer $L$, along with intense peaks at
$[1~0~0]$ and $[0~0~1]$ positions. All of these peaks exhibit the same
temperature dependence, consistent with second-order transitions at 119 and
74~K, respectively, at the two dopings.  The addition of $[0.5~0~L]$ suggests
that the interactions are now ferromagnetic along either $a$ or $b$.  Nuclear
reflections are ordinarily forbidden at $[1~0~0]$ and $[0~0~1]$ because the
lattice is body-centered, and all peaks for $h+k+l$ odd are eliminated by a
$\pi$ phase shift in the neutrons diffracted from the body-centered layer
relative to the other layers. The presence of these reflections implies that
the body-centering is broken at these dopings ({\it i.e.}, that the bilayer at
$z=0.5$ is no longer identical to that at $z=0$ and 1), likely because its
spin orientation is reversed.  The $L$-dependence of these odd-integer peaks;
a comparison of their intensity to allowed nuclear peaks; and the fact that
they follow the same temperature dependence as the magnetic half-integer
peaks, together allow us to conclude that the $[1~0~0]$ and $[0~0~1]$ peaks
are magnetic, not structural, in origin.

Finally, crystals with oxygen contents $6.67 \leq 7-\delta \leq 6.775$ had
peaks at $[0.5~0.5~0.5]$, stronger peaks at $[0.25~0~0]$, and weaker peaks at
some $L$ for all combinations of half and quarter $H$ and $K$; $[0~0~1]$ and
$[1~0~0]$ peaks were absent.  Again all half- and quarter-integer peaks obey a
common second-order temperature dependence, having transitions at 83 and
150~K, respectively, for $7-\delta = 6.67$ and 6.775, but the $[0.5~0.5~0.5]$
and $[0.5~0~0]$ peaks retain a small fraction of their intensity to at least
250~K, with no signs of temperature-dependence above \TN. In SrFeO$_{3-x}$, a
canted antiferromagnetic phase with a propagation vector of $[0.25~0~0]$ (in
the cubic setting) has been identified at high oxygen
contents,\cite{Reehuis2012} and a related phase may be manifested in \SFO,
while the peaks that retain weight above \TN\ suggest an underlying
non-magnetic order.  Identification of the phases in this material at
intermediate dopings and of their propagation vectors will require further
investigation.

\begin{figure}[htb]
\includegraphics[width=\columnwidth]{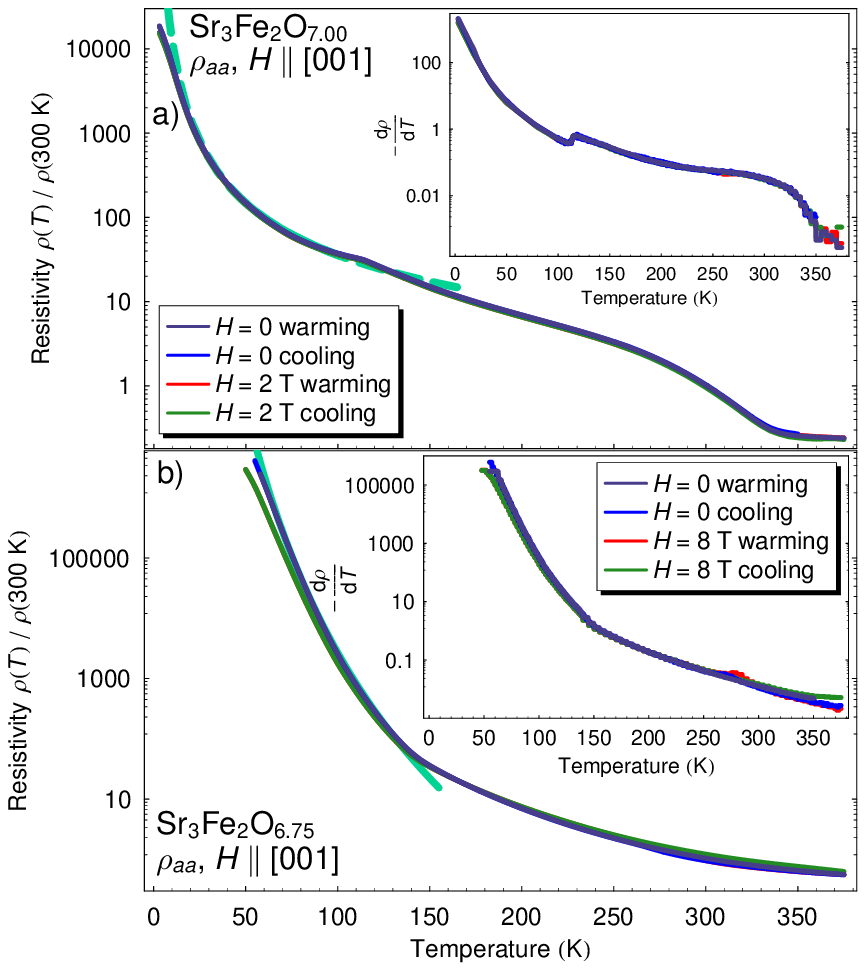}
\caption{\label{fig:rho}(color online) Temperature dependence of the in-plane
  resistivity for a) Sr$_3$Fe$_2$O$_{7.00}$ and b) Sr$_3$Fe$_2$O$_{6.75}$, and
  their derivatives (insets).  Phase transitions are manifested in the
  resistivity as changes in slope; fits of the low temperature behavior to
  three-dimensional variable range hopping are included in light green.  In
  Sr$_3$Fe$_2$O$_{7.00}$, the known magnetic transition at 115~K and
  charge-disproportionation transition \TD\ around 340~K are both visible.}
\end{figure}

Motivated by reports of charge disproportionation in Sr$_3$Fe$_2$O$_{7.00}$,
the in-plane resistivity was measured on several samples in zero field and
again in fields of several Tesla parallel to [001], as displayed in
Fig.~\ref{fig:rho} for two  doping levels;  as mentioned earlier, cracking due
to the intercalation of water prevents displaying these data on an absolute
scale.  Previous x-ray absorption and photoemission work on powders of
SrFeO$_{3-x}$, Sr$_2$FeO$_{4-x}$ and \SFO\ indicated a decrease in bandwidth
as the dimensionality was lowered,\cite{Abbate2004,Abbate2005} and previous
resistivity measurements on \SFO\ indicated this material to be weakly
insulating, with a small excitation gap.\cite{Adler1997,Rozenberg1999}  Our
data are fully consistent with these results.  In contrast to the
three-dimensional cubic perovskite SrFeO$_{3-x}$, which is metallic for high
oxygen contents (becoming insulating at low temperatures as oxygen is
removed), \SFO\ shows insulating low-temperature behavior even at high doping
levels.  At high oxygen contents, the resistivity below the magnetic
transition follows an exponentially-activated $\ln\rho\sim T^{-0.25}$ form
consistent with three-dimensional variable-range hopping, while a
fully-deoxygenated crystal obeys a $\ln\rho\sim 1/T$ form indicative of a
$\sim 270$~meV gap.  There are no indications of hysteresis and minimal field
dependence in fields up to 8~T.

The resistivity data for small $\delta$ exhibit anomalies in the
temperature-dependent resistivity at phase transitions; these are most clearly
visible in the derivative, calculated from a linear interpolation of the
data. For $7-\delta = 7.00$, the resistivity anomalies coincide with the
critical temperatures for charge disproportionation (\TD) and magnetic
ordering extracted from earlier M\"o{\ss}bauer data \cite{Kuzushita2000} and
from the magnetization and neutron scattering data reported above,
respectively.  Although the temperature range of our data above the
charge-disproportionation transition is limited, the substantial resistivity
upturn below this temperature indicates that the weakly-insulating character
is a consequence of the correlation-driven charge disproportionation
instability, rather than disorder. With increasing $\delta$, the charge
disproportionation transition becomes indistinct (Fig.~\ref{fig:rho}).

\section{Conclusion}

Figure~\ref{fig:TN} provides a synopsis of the phase diagram compiled from the
susceptibility, neutron diffraction, and resistivity data presented here. The
insulating end member Sr$_3$Fe$_2$O$_{6.00}$, whose spin system is made up
entirely of Fe$^{3+}$ moments, exhibits commensurate antiferromagnetism with
$T_N \sim 600$ K. This indicates strong unfrustrated superexchange
interactions, analogous to its three-dimensional counterparts
BiFeO$_3$\cite{Sosnowska1996} and NdFeO$_3$,\cite{Bartolome1997} whose N\'eel
temperatures are comparable.

Fully-oxygenated Sr$_3$Fe$_2$O$_{7.00}$, with a spin system comprised entirely
of Fe$^{4+}$ moments, exhibits a sequence of charge-disproportionation and
antiferromagnetic phase transitions in close analogy to CaFeO$_3$, which hosts
a three-dimensional network of Fe$^{4+}$
moments.\cite{Takano1977,Takano1983,Woodward2000} The former transition
generates a small charge excitation gap. The antiferromagnetic state is
characterized by incommensurate long-range order indicative of competing
interactions in the iron oxide bilayers, as a result of which the N\'eel
temperature is much lower than that of Sr$_3$Fe$_2$O$_{6.00}$ despite the
larger moment. The data presented here are not sufficient to establish whether
the magnetism at any doping is helical, but the incommensurate magnetism found
at and near fully-oxygenated Sr$_3$Fe$_2$O$_{7.00}$ is reminiscent of that
found in SrFeO$_{3.00}$,\cite{Takeda1972,Reehuis2012} where helical order was
demonstrated by a detailed refinement of neutron diffraction data. Refining
the magnetic structures identified in \SFO\ is an important task for future
research.  If one of the magnetic phases is indeed found to be helical, it
will constitute one of the few examples of a quasi-two-dimensional helical
magnet.  Since helical spin correlations have been proposed as an explanation
for the magnetism of underdoped La$_{2-x}$Sr$_x$CuO$_4$ between the
superconducting and antiferromagnetic phases,\cite{Luscher2007} \SFO\ may turn
out to be a model system for such magnetism, offering crucial context for the
cuprates.

For intermediate oxygen contents, different forms of magnetic order are
observed, with transition temperatures that do not approach zero as previously
proposed. This indicates that Fe$^{3+}$-Fe$^{4+}$ interactions and oxygen
vacancies frustrate the exchange interactions in a non-trivial manner. Similar
observations have been made in  SrFeO$_{3-x}$.\cite{Reehuis2012} The nature
and origin of the frustration, the magnetic propagation vectors, and the order
associated with the susceptibility anomalies in Fig.~\ref{fig:TN}, are
interesting subjects for future research.

\begin{acknowledgments}

We thank O. P. Sushkov and G. Khaliullin for helpful discussions, the members
of the MPI-FKF Crystal Growth group and Jansen's Department for assistance,
and the German Science Foundation (DFG) for financial support under
collaborative grant No.\ SFB/TRR~80.  A portion of this work was carried out
at the Swiss spallation neutron source SINQ, Paul-Scherrer-Institut, Villigen,
Switzerland.

\end{acknowledgments}

\bibliography{327_doping}

\end{document}